%% file: main.tex
\PassOptionsToPackage{numbers,sort&compress}{natbib}
\documentclass[preprint,12pt, a4paper]{elsarticle}
\usepackage[dvipsnames,table,xcdraw]{xcolor}

\newcommand{\reviewerOne}[1]{{#1}}
\newcommand{\reviewerTwo}[1]{{#1}}

\newcommand{\reviewerThree}[1]{{#1}}
\newcommand{\reviewerFour}[1]{{#1}}

\title{SynthPix: A lightspeed PIV \reviewerOne{image} generator}
\author[eth]{Antonio Terpin\fnref{corresponding,eqcontrib}}
\author[eth]{Alan Bonomi\fnref{eqcontrib}}
\author[eth]{Francesco Banelli}
\author[eth]{Raffaello D'Andrea}
\address[eth]{\reviewerThree{Institute for Dynamic Systems and Control, ETH Z\"urich, Z\"urich, Switzerland}}
\fntext[corresponding]{Corresponding author.}
\fntext[eqcontrib]{Equal contribution.}
\journal{SoftwareX}
\usepackage{graphicx}

\usepackage{amsmath,amssymb,mathtools,bm,amsthm,bbm,dsfont,stmaryrd,aligned-overset,physics}

\theoremstyle{remark}
\newtheorem{remark}{Remark}[section]

\usepackage{booktabs}
\usepackage{multirow}
\definecolor{lightgray}{gray}{0.97}
\rowcolors{2}{white}{lightgray}
\usepackage{adjustbox}
\usepackage{tcolorbox}
\usepackage{enumitem}
\usepackage{mdframed}
\usepackage{amsthm}

\usepackage{wrapfig}
\usepackage{subcaption}

\usepackage{tikz,pgfplots}
\usetikzlibrary{patterns,hobby,calc,intersections}
\pgfplotsset{compat=1.18}
\usepgfplotslibrary{statistics,fillbetween}
\usepackage{pgfplotstable}

\usepackage[acronym]{glossaries}
\glsdisablehyper
\newacronym{ccd}{CCD}{Charge Coupled Device}
\newacronym{piv}{PIV}{Particle Image Velocimetry}
\newacronym{epe}{EPE}{end-point error}

\usepackage{pifont}
\renewcommand{\checkmark}{\textcolor{black}{\ding{51}}}
\newcommand{\notcheck}{\textcolor{black}{\ding{55}}}
\newcommand{\maybecheck}{\textcolor{black}{\textemdash}}

\usepackage{microtype}

\usepackage{hyperref}
\usepackage[capitalize,noabbrev]{cleveref}
\usepackage{url}

\usepackage{amssymb}

\usepackage{listings}
\usepackage{tcolorbox}
\tcbuselibrary{listings}

\crefname{lstlisting}{\reviewerThree{Listing}}{\reviewerThree{Listings}}
\Crefname{lstlisting}{\reviewerThree{Listing}}{\reviewerThree{Listings}}

\definecolor{codebg}{RGB}{247,247,247}

\lstdefinestyle{pythoncode}{
  language=Python,
  backgroundcolor=\color{codebg},
  frame=none,
  basicstyle=\ttfamily\small,
  keywordstyle=\bfseries\color{RoyalBlue},
  commentstyle=\itshape\color{ForestGreen},
  stringstyle=\color{BrickRed},
  showstringspaces=false,
  columns=fullflexible,
  keepspaces=true,
  breaklines=true,
  breakatwhitespace=true,
}

\begin{document}
\begin{frontmatter}
\begin{abstract}
We describe SynthPix, a synthetic image generator for Particle Image Velocimetry (PIV) with a focus on performance and parallelism on accelerators, implemented in JAX.
\reviewerOne{
SynthPix produces PIV image pairs from prescribed flow fields while exposing a configuration interface aligned with common PIV imaging and acquisition parameters (e.g., seeding density, particle image size, illumination nonuniformity, noise, blur, and timing). In contrast to offline dataset generation workflows, SynthPix is built to stream images on-the-fly directly into learning and benchmarking pipelines, enabling data-hungry methods and closed-loop procedures---such as adaptive sampling and acquisition/parameter co-design---without prohibitive storage and input-output costs. We demonstrate that SynthPix is compatible with a broad range of application scenarios, including controlled laboratory \reviewerFour{experiments and} riverine image velocimetry, and supports rapid sweeps over nuisance factors for systematic robustness evaluation.}
\reviewerOne{SynthPix is a tool that supports the flow quantification community} and in this paper we describe the main ideas behind the software package.
\end{abstract}
\end{frontmatter}

\setcounter{topnumber}{5}
\setcounter{bottomnumber}{5}
\setcounter{totalnumber}{10}
\renewcommand{\topfraction}{0.95}
\renewcommand{\bottomfraction}{0.95}
\renewcommand{\textfraction}{0.05}
\renewcommand{\floatpagefraction}{0.8}

\clearpage
\section*{Code metadata}
\begin{table}[htb!]
\begin{tabular}{|l|p{6cm}|p{6cm}|}
\hline
\textbf{Nr.} & \textbf{Code metadata description} & \textbf{Metadata} \\
\hline
C1 & Current code version & v0.1.0 \\
\hline
C2 & Permanent link to code/repository used for this code version & \url{https://github.com/antonioterpin/synthpix} \\
\hline
C3  & Permanent link to Reproducible Capsule & \url{https://github.com/antonioterpin/synthpix/blob/main/src/main.py}\\
\hline
C4 & Legal Code License   & MIT License. \\
\hline
C5 & Code versioning system used & git\\
\hline
C6 & Software code languages, tools, and services used & Python\\
\hline
C7 & Compilation requirements, operating environments \& dependencies & Python 3.10+, Jax 0.6.2+, tqdm 4.67.1, h5py 3.13.0+, ruamel.yaml 0.18.10+, robo-goggles 0.1.7+\\
\hline
C8 & If available Link to developer documentation/manual &\url{https://github.com/antonioterpin/synthpix} \\
\hline
C9 & Support email for questions & aterpin@ethz.ch\\
\hline
\end{tabular}
\caption{Code metadata}
\label{codeMetadata} 
\end{table}

\input{figure_1}
\clearpage
\section{Motivation and significance}
\label{sec:intro}

\reviewerTwo{\gls*{piv} \cite{Grant1997,Adrian1991,Adrian2005,willert2007particle,westerweel2013particle,Scharnowski2020} and related image-based flow-quantification methods \cite{Liu2008,corpetti2006fluid,zhong2017optical} are among the central modalities for measuring flow fields across laboratory \cite{Adrian1991,Adrian2005}, industrial \cite{willert2007particle,tropea2007springer}, microfluidic \cite{Santiago1998,Meinhart1999}, \reviewerOne{and environmental settings \cite{Muste2008,Fujita2007,Perks2020,legleiter2024framework}.}}
\reviewerTwo{As application domains have broadened, so too have the underlying estimation methodologies. Correlation-based \gls*{piv} remains the backbone of the field \cite{willert1991digital,Keane1992,scarano2001iterativeimage,wang2020globally,lee2024surrogate}, while optical-flow-based methods provide an important complementary family of approaches \cite{horn1981determining,Liu2008,corpetti2006fluid,zhong2017optical,kroeger_fast_2016,pimienta_towards_2024}.}
\reviewerOne{These applications are not limited to laboratory settings: similar image-based workflows arise in field deployments using bank-mounted, bridge-mounted, and airborne imaging platforms for natural flows, where LSPIV and STIV are established tools for non-intrusive surface-velocity estimation \cite{Muste2008,Fujita2007,LeCoz2010,Perks2020,legleiter2024framework}.}

\reviewerTwo{Community benchmark efforts \reviewerFour{show that} rigorous method comparison \reviewerFour{remains} a central part of \gls*{piv} methodology. The International \gls*{piv} Challenge, in particular, helped \reviewerFour{standardize} comparison on shared test cases with common evaluation procedures and reference solutions as a community practice \cite{Stanislas2003,Stanislas2005,Stanislas2008,kahler_main_2016}. Such benchmark datasets are highly valuable, but they are also necessarily finite and task-specific: they are assembled for particular modalities and operating conditions, and therefore cannot span the full range of particle densities, image diameters, noise levels, blur, illumination non-uniformities, displacement gradients, and out-of-plane losses that affect estimator behavior in practice.}
Synthetic generators make it possible to isolate, under reproducible conditions, the effects of particle image diameter, seeding density, noise, blur, illumination non-uniformity, displacement gradients, and out-of-plane loss, from early synthetic-image efforts for \gls*{piv} assessment and benchmarking \cite{Okamoto2000,Stanislas2003,lecordier2004europiv} to more recent benchmarking-oriented software and simulation frameworks \cite{mendes2020piv,Rajendran2019}.
This role is important not only for controlled benchmarking, sensitivity analysis, uncertainty studies, and robustness evaluation, but also for learning-based \gls*{piv}, where simple neural networks \cite{hassan1997new,chen1998artificial,Rabault_2017} have evolved into data-hungrier architectures \cite{cai2019dense,manickathan2022kinematic,lagemann2021deep,REDDY2025120205,zhu2025pivflowdiffusertransferlearningbaseddenoisingdiffusionmodels}.

\reviewerOne{However, offline dataset generation still does not cover the main workflows targeted by SynthPix because of scale and rigidity. Modern simulation collections already span multi-terabyte regimes before any \gls*{piv} rendering, as illustrated by \emph{The Well} (15~terabytes across 16 datasets) \cite{ohana2024the} and the \emph{Johns Hopkins Turbulence Database} (well over 200~terabytes) \cite{jhtdb}. Even using only \(100\) flow fields from such collections, together with five values each for seeding density, particle diameter, noise level, illumination level, and \(\Delta t\), already yields hundreds of thousands of conditions. \reviewerFour{At \(1024\times1024\) resolution, storing a single uint8 image pair for each condition already requires nearly one terabyte, and storing \(100\) pairs per condition requires tens of terabytes.} At that scale, generation, storage, transfer, and repeated reading during training become \reviewerFour{bottlenecks}. Offline generation is also rigid: changing the flow source or rendering parameters requires regenerating samples, which is problematic for settings such as hard-example mining, robustness sweeps, and acquisition-parameter co-design that rely on an evolving sampling distribution.}

\begin{mdframed}[hidealllines=true,backgroundcolor=blue!5]
\vspace{-0.25cm}
\paragraph{Contributions} 
\reviewerOne{We present SynthPix, an open-source Python package for synthetic \gls*{piv} image generation, implemented in JAX and designed for hardware accelerators and direct integration into learning and benchmarking pipelines. Unlike existing generators aimed primarily at offline image creation, SynthPix enables high-throughput batched generation on GPUs and streams images on-the-fly into downstream estimators and training loops. It retains the flexibility expected from established \gls*{piv} generators, exposing physically meaningful imaging and acquisition parameters while supporting diverse flow-field sources through a unified interface. Empirically, SynthPix delivers image-pair throughput orders of magnitude beyond existing alternatives, making large-scale robustness studies and adaptive workflows practical. With documentation, tests, and an extensible design, it provides a reusable foundation for reproducible synthetic \gls*{piv} data generation. See \cref{fig:cover} for an overview.}
\end{mdframed}

\section{Software description} 
\label{sec:design-choices}
We developed SynthPix in parallel with related projects \cite{banelli2025admm,terpin2025flow,terpin2025ff,grontas2025pinetoptimizinghardconstrainedneural}, which helped us design a flexible, easy-to-integrate data-generation pipeline.
In this and the following section, we outline our main design choices; for a comprehensive explanation of the software and detailed tutorials, see \reviewerThree{\cref{codeMetadata}}.

\paragraph{Usage}
\cref{lst:api} shows the simple API we designed for SynthPix.

\begin{lstlisting}[
        style=pythoncode,
        caption={\reviewerThree{Using SynthPix to instantiate the image generator, get a new batch of \gls*{piv} images, the corresponding ground-truth flows and the sampled parameters.}},
        label={lst:api},
        float,
        captionpos=b,
    ]
    # pip install synthpix
    import synthpix

    sampler = synthpix.make("/path/to/config.yaml")

    batch = next(sampler)
    images1 = batch.images1
    images2 = batch.images2
    flows = batch.flow_fields
    params = batch.params
    
    print(f"{images1.shape=}")
    print(f"{images2.shape=}")
    print(f"{flows.shape=}")

    # We can access the sampled generation parameters, for instance:
    print(f"{params.seeding_densities.shape=}")

    # pseudo-code in a loop:
    while my_cond:
        batch = next(sampler)
        images1 = batch.images1
        images2 = batch.images2
        flows = batch.flow_fields
        flows_estimated = my_estimator(images1, images2)
        print(f"EPE = {epe(flows_estimated, flows)}")
\end{lstlisting}

\paragraph{GPU enabled} The user can specify in the configuration file which device to use (e.g., ``cpu'', ``cuda:0'', or \reviewerOne{multiple GPUs}).

\paragraph{Configuration file}
The configuration file specifies all options SynthPix needs to generate synthetic \gls*{piv} images. SynthPix allows the user to specify 
the image shape (e.g., $1024\times1024$), the batch size (e.g., $B = 256$), how many different flow fields to use for each batch (e.g., $F = 64$, so that $B / F = 256/64 = 4$ image pairs are generated from each flow field), and for how many consecutive batches \reviewerOne{a loaded flow field is reused} (e.g., $1$).
SynthPix also provides several configurable \reviewerOne{generation and noise parameters}.

\paragraph{Data loaders and pre-fetching}
\reviewerTwo{SynthPix decouples image generation from the origin and storage format of the underlying flow field through a unified flow-provider interface. In practice, flow fields specified in the configuration are loaded and transferred to the target device for rendering, with optional pre-fetching to overlap data movement and computation.}

\reviewerTwo{The current implementation already supports several common representations, including \texttt{.hdf5} files from the Johns Hopkins Turbulence Database \cite{Li2008,Perlman2007}, \texttt{.mat} files as in \cite{JassalSchmidt2024}, and generic \texttt{.npy} arrays storing tuples of the form $(x,y,u(x,y),v(x,y))$. Existing \gls*{piv} images can also be loaded alongside the flow field, enabling direct use of datasets that already provide paired images and velocities \cite{cai2019dense,wei2025velocity}.}
\reviewerTwo{More importantly, this interface is not tied to a fixed catalogue of datasets: flow fields may also be supplied procedurally through a Python function, which makes it straightforward to integrate analytically defined or randomly generated flows \cite{10.1115/1.1369598,manickathan2022kinematic}.
This includes, for example, web-accessible turbulence repositories and DNS databases for canonical turbulent flows \cite{Perlman2007,Li2008,graham2016web}, broader simulation libraries spanning multiple physical regimes and coupled PDE systems \cite{ohana2024the,takamoto_pdebench_2022}, specialized CFD datasets for turbulent wakes and related engineering flows \cite{wakeset,towne_database_2023,luo_cfdbench_2023}, and river-image-velocimetry frameworks and benchmark datasets for algorithm evaluation under realistic hydraulic and imaging conditions \cite{legleiter2024framework,patalano_rectification_2017,bodart_synthetic_2022,Perks2020}.}

\paragraph{Mathematical model for the contribution of the single particle}
We model the contribution of the single particle to the pixel intensity $I(x, y)$ in the image as
\begin{equation}
\label{eq:single-contrib}
    I_0 \exp\left(-\frac{1}{2(1-\rho^2)}\left(\frac{(x - x_0)^2}{\sigma_x^2} - 2\rho\frac{(x - x_0)(y - y_0)}{\sigma_x\sigma_y} + \frac{(y - y_0)^2}{\sigma_y^2}\right)
\right),
\end{equation}
where $I_0$ is the peak intensity, \reviewerOne{$(x_0, y_0)$ is the (continuous) particle position,} and the parameters $\sigma_x, \sigma_y, \rho$ determine the shape of the contribution. This is the classical model used for rendering synthetic \gls*{piv} images, as presented in \cite{willert2007particle}, with the difference that allowing $\rho > 0$ also permits axis-misaligned particles. All the parameters are sampled uniformly at random from a configurable interval for each particle in the first image. For the particles in the second image, we sample the variation of each parameter for each particle from a zero-mean Gaussian with specified variance.

\begin{remark}\label{remark:out-of-focus}
    A different approach for calculating $I_0$, which we defer to future releases, is to consider the particles \reviewerOne{in three}-dimensional space, apply a three-dimensional flow, and then compute
    $
        I_0(z) = q\exp\left(-\frac{1}{\sqrt{2\pi}}\left\vert\frac{2Z^2}{\Delta Z_0^2}\right\vert^s\right),
    $
    where the \reviewerOne{particle} scattering efficiency $q$, the thickness of the laser sheet $\Delta Z_0$, and the shape factor $s$ are configurable parameters; see \cite{willert2007particle}.
\end{remark}

\begin{remark}\label{remark:camera-model}
\reviewerTwo{The current version assumes an orthogonal imaging geometry and a simplified sensor model.
    In particular, we do not yet model non-orthogonal projection effects \cite{lecordier2004europiv} or detector-specific sampling properties such as finite pixel fill factor.
    We note that such effects depend on the sensing technology and imaging hardware, and should therefore not be interpreted as uniformly applying across camera types: for example, CCD- and CMOS-based \reviewerFour{systems differ} in the extent to which these effects are relevant in practice.
    These imaging- and sensor-modeling extensions are deferred to future releases.}
\end{remark}

Finally, SynthPix-generated images are optionally adjusted according to a user-specified intensity histogram.

\paragraph{Sampling the particles}
\reviewerOne{For the first image, we allocate particles according to the maximum configured particles-per-pixel value \texttt{ppp\_max} and sample positions for $H\times W\times \texttt{ppp\_max}$ candidate particles over the image domain. We then sample the actual \texttt{ppp} uniformly from the configured range and mask the excess particles by setting their intensities to zero.}
For the second image, we move the particles according to the mapping $(x, y) \mapsto (x + u(x, y), y + v(x, y))$, where $(x, y)$ are the particle coordinates and $(u(\cdot, \cdot), v(\cdot, \cdot))$ are the flow field coordinates; see \cref{fig:cover}.

\paragraph{Considerations on the aggregation of the \reviewerOne{particle} contributions}
We sum the contributions of all the particles to obtain the resulting image. This superposition is commonly adopted in the literature, and relies both on a low number of particles per pixel (\texttt{ppp}) and on the absence of interference phenomena \cite{lecordier2004europiv}. \reviewerOne{Modeling these effects represents a meaningful extension of SynthPix and will be part of a future release.} In the current version, to mitigate this, we allow the user to select a probability of hiding a particle from each image. \reviewerOne{That is, a particle may be rendered in the first image but not in the second, and vice versa.} To exploit parallelism on accelerators with JAX, we keep the number of particles constant by setting the intensity of the hidden particles to zero.

\paragraph{Efficient rasterization}
To efficiently sum the contributions of the particles, we first generate, in parallel, a batch of Gaussian kernels, one for each particle, with patch sizes of $3 \max(\{d_i\}) + 1$, where $d_i$ are the diameters of the particles. These patches are generated according to the selected mathematical model for the single particles. Then, we sum these kernels to obtain the final image.

\section{Illustrative examples}
\label{sec:results}
We compare SynthPix to existing alternatives and analyze its \reviewerOne{performance} in different scenarios. The empirical data \reviewerOne{is} collected on an Ubuntu 22.04 machine equipped with an AMD Ryzen Threadripper PRO 5995WX processor and four Nvidia RTX 4090 GPUs. The number of GPUs used and the exact experimental setup is detailed for each experiment below.

\subsection{Comparisons to related software}
There are several alternatives for synthetic \gls*{piv} \reviewerOne{image} generation; see, for instance, \cite{lecordier2004europiv,stamhuis2014pivlab,Thielicke2021,probst_synpivimage_2024,willert2007particle,mendes2020piv}. However, the requirements of modern deep-learning approaches call for a pipeline with focus on parallelism on accelerators, compatible with existing machine learning infrastructure and that facilitates batched generation of large quantities of data. Moreover, to facilitate (i) the sharing of algorithmic advancements, (ii) the benchmarking with the state-of-the-art, and (iii) the reproducibility of the proposed pipelines, we need an easy-to-install, easy-to-use, and easy-to-contribute-to software following open-source software engineering best practices. Existing methods do not satisfy these conditions: in fact, most existing work on deep-learning for fluid flow quantification uses their own dataset generator \cite{cai2019particle,wei2025velocity,cai2019dense,manickathan2022kinematic,gao2021robust,lagemann2021deep,Rabault_2017,lee2017piv}.
\reviewerOne{SynthPix has} an edge over other packages in terms of usability, documentation, testing, and integrability with modern approaches; see \cref{tab:comparisons}. \reviewerOne{Moreover, its throughput vastly outperforms the existing alternatives.}

\begin{table}[]
    \centering
    \begin{adjustbox}{width=1\textwidth,center}
    \begin{tabular}{l l l l l l}
& SynthPix & Synpivimage \cite{probst_synpivimage_2024,willert2007particle} & PIV image gen. \cite{mendes2020piv} & PIVlab \cite{stamhuis2014pivlab,Thielicke2021} \\\hline
        Language/Backend & Python/JAX & Python/Numpy & MATLAB & MATLAB \\
        Integrates with PyTorch/JAX & \checkmark & \maybecheck & \notcheck & \notcheck \\
        Allows batched generation & \checkmark & \notcheck & \notcheck & \notcheck \\
        User guide / Tutorials & \checkmark / \checkmark & \checkmark / \checkmark & \checkmark / \checkmark & \checkmark / \checkmark \\
        API Documentation & \checkmark & \checkmark & \notcheck & \notcheck\\
        Benchmarks & \checkmark & \notcheck & \notcheck & \notcheck \\
        Ready to use data & \checkmark & \notcheck & \checkmark & \checkmark \\
        Test coverage & \checkmark & \notcheck & \notcheck & \notcheck \\
        Issue / PR template & \checkmark & \notcheck & \notcheck & \checkmark \\\hline
    \end{tabular}
    \end{adjustbox}
    \caption{Comparison of SynthPix to the available alternatives. Marks: \checkmark{} means that the feature is deemed fully provided, \notcheck{} if it is not, \maybecheck{} if it is deemed possible with a (possibly inefficient) workaround.}
    \label{tab:comparisons}
\end{table}

\subsubsection{Throughput}
\paragraph{Experimental setup} We compare the throughput of each method by generating a number of $512\times 512$ image pairs. Because the methods vastly differ in the generation speed and interface, we collect $100$ image pairs for each baseline \reviewerOne{\cite{probst_synpivimage_2024,willert2007particle,mendes2020piv,stamhuis2014pivlab,Thielicke2021}} and for our method, instead, $100\text{'}000$. Moreover, since our method is the only one that \reviewerOne{is batched and can} exploit a hardware accelerator, we collect data both on CPU and (a single, see \cref{sec:results:scaling-laws} for an ablation on the number of GPUs) GPU and with batch sizes of $1$ and $256$.

\paragraph{Results} The results are on the bottom left of \cref{fig:cover}. In every setting, SynthPix outperforms all other methods by orders of magnitude. In particular, the results show the benefit of exploiting hardware acceleration.
We juxtapose two image pairs generated with PIVlab \cite{stamhuis2014pivlab,Thielicke2021} and SynthPix in \cref{fig:images-generated}.
\input{figure_2}

\subsection{\reviewerTwo{Quality of the generated images}}
\label{sec:results:quality}
\reviewerTwo{We assess the quality of SynthPix-generated images through their usefulness for downstream \gls*{piv} \reviewerFour{estimation.} Across the following experiments, we compare estimator behavior on original data and on images generated with SynthPix \reviewerFour{to assess if} the synthetic images preserve the information relevant to benchmarking, analysis, and training. As the evaluation metric, we use the \gls*{epe}, defined as the mean pixelwise $\ell_2$ distance between an estimated flow field and a reference flow field, averaged over the pixels $(x,y)$ in the image domain $I$,}
\reviewerOne{
\begin{equation}
\label{eq:epe_def}
\mathrm{EPE}(u,v) \coloneqq \frac{1}{|I|}
\sum_{(x,y)\in I} \left\|u(x,y)-v(x,y)\right\|_2.
\end{equation}}

\subsubsection{\reviewerTwo{Estimator self-consistency on experimental PIV data}}

\paragraph{Experimental setup}
\reviewerTwo{We next consider experimental data from case A of the $2^{\mathrm{nd}}$ PIV Challenge~\cite{Stanislas2003}. Since no ground-truth flow is available for the experimental data, we assess \reviewerOne{re-rendering self-consistency}. For each original image pair \(I_i^{\mathrm{orig}}\), we first estimate a flow field \(u_i\) with JPIV, generate a synthetic image pair \(I_i^{\mathrm{syn}}\) from \(u_i\) using SynthPix, and then re-estimate the flow \(\tilde u_i\) from the synthetic images with JPIV. We report the average self-consistency error
\begin{equation}
\label{eq:self_consistency}
\mathcal{C}_{\mathrm{avg}}
\coloneqq
\mathrm{average}(\mathrm{EPE}(\tilde u_i,u_i))
\quad\text{and}\quad
\mathcal{C}_{\mathrm{median}}
\coloneqq
\mathrm{median}(\mathrm{EPE}(\tilde u_i,u_i)),
\end{equation}
where average and median are taken over all the image pairs. Smaller values of \(\mathcal{C}_{\mathrm{avg}}, \mathcal{C}_{\mathrm{median}}\) indicate that SynthPix better preserves the flow information recovered by the estimator from the original experimental data.}

\paragraph{Results}
\reviewerTwo{The resulting self-consistency errors are \(\mathcal{C}_{\mathrm{avg}}=0.536\) and \(\mathcal{C}_{\mathrm{median}}=0.368\), indicating that re-estimating the flow from SynthPix-generated image pairs yields results close to those obtained from the original experimental images \reviewerFour{and is adequate for practical applications.}}

\subsubsection{\reviewerOne{Benchmark-error prediction on synthetic reference data}}

\paragraph{Experimental setup}
We use the synthetic \gls*{piv} dataset \cite{cai2019dense,wei2025velocity}, which provides image pairs together with ground-truth flow fields. For each image pair, we evaluate the DIS optical-flow estimator \cite{kroeger_fast_2016,teed_raft_2020} via Flow Gym \cite{terpin2025flow} on both the original images and on corresponding images generated with SynthPix, and compare the resulting \gls*{epe} values with respect to the same ground-truth flow field. To match the heterogeneous rendering conditions present in the original dataset, we tune the SynthPix generation parameters for each case to reproduce the visual characteristics of the corresponding reference images as closely as possible.
\reviewerOne{To quantify how well evaluation on SynthPix predicts evaluation on the established benchmarks, we report the \emph{relative EPE \eqref{eq:epe_def} discrepancy}}
\begin{equation}
\label{eq:rel_epe_discrepancy}
\mathcal{L}(u_1,u_2) \coloneqq 
\frac{\left|\mathrm{EPE}(u_1,\hat u)-\mathrm{EPE}(u_2,\hat u)\right|}{\mathrm{EPE}(u_1,\hat u)},
\end{equation}
where $u_1$ and $u_2$ denote the flows estimated from the original and SynthPix-generated images, respectively, and $\hat u$ denotes the ground-truth flow. \reviewerOne{A value of $\mathcal{L}(u_1,u_2) = 0$ indicates that both evaluations yield identical \gls*{epe} \eqref{eq:epe_def}, while larger values indicate a larger deviation of SynthPix-based evaluation from the original benchmark.}

\input{figure_4}

\paragraph{Results} Over all image pairs, we obtain a relative discrepancy on the \gls*{epe} of at most $0.005$, averaging at $0.001$, confirming the quality of the synthetic images generated with SynthPix. We also showcase, for different flow types in \cite{cai2019dense}, a sample of images from \cite{cai2019dense} and from SynthPix in \cref{fig:accuracy}.

\subsubsection{\reviewerTwo{Training learning-based estimators with SynthPix-generated data}}

\paragraph{Experimental setup}
\reviewerTwo{Finally, we evaluate SynthPix as a tool for generating training data for learning-based \gls*{piv} estimation. We use flow fields derived from the PIV dataset to generate a substantially larger synthetic training set with SynthPix, train a RAFT-based estimator from Flow Gym \cite{terpin2025flow} for three hours and fixed learning rate $\mathrm{lr}=0.0001$ on these generated image pairs, and test it on the original images. Performance is reported as the average \gls*{epe} \eqref{eq:epe_def} on the original test set.}

\paragraph{Results}
\reviewerTwo{Training on the enlarged SynthPix-generated dataset yields an average test \gls*{epe} of $0.065$ on the original images. Despite being trained on synthetic renderings rather than on the original images themselves, the model generalizes to the original test set with competitive accuracy, supporting the use of SynthPix for large-scale training data generation.}

\subsection{Ablations}
\label{sec:results:scaling-laws}
\input{figure_5}

In this section, we assess the performance impact of the generation parameters.
\paragraph{Experimental setup} We perform ablations on the number of GPUs, image size, seeding density, maximum \reviewerOne{particle diameter}, minimum seeding density in the batch with fixed maximum, batch size, flow fields per batch, and batches generated with the same flow field.
For each ablation study, we report the throughput in image pairs per second. The parameters that are not under study are fixed. In particular, we use a single GPU and consider an image size of $512\times512$, a seeding density of $0.06$, \reviewerOne{particle diameters} sampled uniformly in $[0.8, 1.2]$, a batch size of $64$, and $1$ flow field per batch, kept for all the batches. The statistics are computed over $1000$ batches.
\paragraph{Results} We collect the results in \cref{fig:ablations-throughput} and make the following observations:
\begin{itemize}[leftmargin=*]
    \item The throughput scales linearly with increasing number of GPUs used, as one would expect from a software exploiting parallelism on accelerators.
    \item The throughput decreases with increasing image size, but remains high even for very large images, enabling the generation of high-resolution datasets.
    \item The throughput mildly decreases with increasing seeding density, approximately in a linear relationship. The throughput is not affected, instead, by the spread between the minimum and the maximum seeding density. 
    \item The throughput decreases with increasing maximum \reviewerOne{particle diameter}, but the trend resembles a staircase: the maximum \reviewerOne{particle diameter} affects in discrete steps the Gaussian kernel patch size.
    \item The throughput increases substantially with the batch size initially, and then only mildly. This is not unexpected, since the increase in batch size amortizes the initial overheads of the accelerators, but these benefits are substantial only when moving from a single batch \reviewerOne{to multiple}.
    \item The throughput is not affected by the number of flow fields used during generation, \reviewerOne{consistent} with the parallelization over the flow fields.
    \item The throughput increases only slightly with the number of batches with the same flow field. Future releases will load the data directly to the device.
\end{itemize}

\section{Impact}
\label{sec:impact}
SynthPix empowers researchers with a high-throughput pipeline to generate realistic and high-resolution \gls*{piv} images compatible with modern deep-learning frameworks. SynthPix is written in JAX, enabling efficient large-scale training on hardware accelerators. With SynthPix, researchers can now push the limits of the use of synthetic data in the development of flow quantification methods and deploy more data-hungry algorithms.
SynthPix has already supported multiple research works in adaptive PIV tuning~\cite{banelli2025admm}, 
large-scale flow quantification~\cite{terpin2025flow}, assessment of hard-constrained neural networks \cite{grontas2025pinetoptimizinghardconstrainedneural}, and real-time feedback control~\cite{terpin2025ff}. 
Beyond these applications, SynthPix standardizes the generation of synthetic PIV datasets, removing the need for each study to ``reinvent the wheel''. 
\reviewerOne{Its design and interoperability promote reproducibility} and adoption across the machine-learning and fluid-dynamics communities, 
laying the foundation for new advances in data-driven flow quantification.

\section{Conclusions}
In this paper, we describe SynthPix, a synthetic image generator for \gls*{piv} implemented in JAX, designed to maximize throughput via batching and parallelism on accelerators while retaining the configuration flexibility of established generators. Sections~\ref{sec:results}--\ref{sec:impact} show that SynthPix enables large-scale generation of high-resolution image pairs and supports workflows where data generation must keep pace with modern learning-based training and benchmarking pipelines.
In the future, we plan to extend SynthPix by:
\begin{itemize}[leftmargin=*]
    \item Implementing the second approach for computing $I_0$ described in \Cref{remark:out-of-focus} and non-orthogonal configurations for the camera model (see \Cref{remark:camera-model})
    \item Enabling the synthesis of \gls*{piv} images from multiple cameras in a volumetric setting, as needed for tomographic PIV and PTV pipelines \cite{Schanz2016}.
    \item Reusing \reviewerOne{particle} positions across consecutive frames, i.e., applying the flow for a second, possibly different, $\Delta t'$ to the ``next particles positions'' in \cref{fig:cover} to obtain the ``initial \reviewerOne{particle} positions'' for the successive batch.
\end{itemize}
Ultimately, we will prioritize the needs of the community as expressed via open issues and pull requests.

\section*{Acknowledgement}
The authors would like to acknowledge the editor and the anonymous reviewers for their valuable and constructive feedback. We believe that the review process substantially strengthened the manuscript and improved the presentation of our contribution.

\bibliographystyle{elsarticle-num} 
\bibliography{references}

\end{document}

%% file: figure_1.tex
\begin{figure}[htb]
    \centering
    \begin{minipage}{\linewidth}
    \centering
        \includegraphics[width=\linewidth]{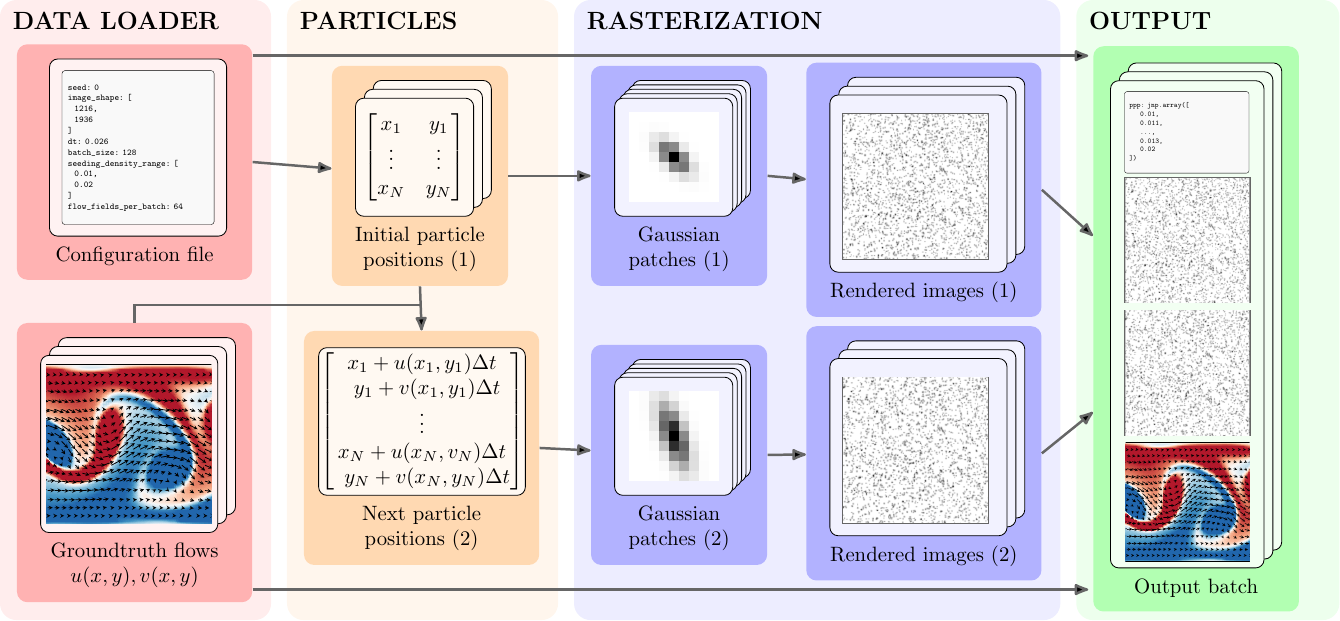}
        \vspace{.25cm}
    \end{minipage}
    
    \vspace{.25cm}
    
    \begin{minipage}{.49\linewidth}
    \centering
    \begin{adjustbox}{max width=\linewidth}
    \begin{tikzpicture}
      \begin{semilogyaxis}[
        log origin=infty,
        width=7.5cm,
        height=5.5cm,
        ylabel={Image pairs per second},
        ytick={1,10,100,1000,10000},
        symbolic x coords={Synpivimage,PIV image generator,PIVlab,Ours},
        xtick=data,
        axis on top,
        enlarge x limits=0.2,
        xticklabels={
          {\cite{probst_synpivimage_2024}},
          {\cite{mendes2020piv}},
          {\cite{stamhuis2014pivlab}},
          {Ours}
        },
        legend columns=1,
        legend style={at={(0.02,0.98)}, anchor=north west},
        legend cell align = {left},
        unbounded coords=discard,
      ]
    
        \addplot+[
          ybar,
          bar width=25pt,
          bar shift=0pt,
          draw=black,
          fill=white,
          pattern=crosshatch,
          pattern color=black,
        ] coordinates {
          (Synpivimage,          0.89)
          (PIV image generator,  1.19)
          (PIVlab,               5.97)
          (Ours,                 55)
        };
    
        \addplot+[
          ybar,
          bar width=25pt,
          bar shift=0pt,
          draw=black,
          fill=white,
          pattern=north west lines,
          pattern color=black,
        ] coordinates {
          (Synpivimage,          0)
          (PIV image generator,  0)
          (PIVlab,               0)
          (Ours,              10477)
        };
    
        \addplot+[
          ybar,
          bar width=25pt,
          bar shift=0pt,
          draw=black,
          pattern color=black,
        ] coordinates {
          (Synpivimage,          0)
          (PIV image generator,  0)
          (PIVlab,               0)
          (Ours,              22538)
        };
    
        \legend{CPU $B = 1$,GPU $B = 1$,GPU $B = 256$}
      \end{semilogyaxis}
    \end{tikzpicture}
    \end{adjustbox}
    \end{minipage}
    \begin{minipage}{.49\linewidth}
    \centering
    \vspace{-.75cm}
    \begin{adjustbox}{max width=\linewidth}
    \begin{tabular}{l l}
    & \bf SynthPix
    \\\hline
        Language/Backend & Python/JAX \\
        Integrates with PyTorch/JAX & \checkmark \\
        Allows batched generation & \checkmark \\
        User guide / Tutorials & \checkmark / \checkmark\\
        API Documentation & \checkmark\\
        Benchmark & \checkmark\\
        Ready to use data & \checkmark\\
        Test coverage & \checkmark\\
        Type checking & \checkmark\\
        Issue / PR template & \checkmark/\checkmark\\\hline
    \end{tabular}
    \end{adjustbox}
    \end{minipage}
    \caption{The SynthPix pipeline (top) follows established PIV image-generation techniques but is optimized for performance on hardware accelerators.
    Below, we compare SynthPix’s throughput against existing synthetic particle-image generators (B: batch size; CPU/GPU: execution device) and summarize the key characteristics of the SynthPix codebase;
    see \cref{sec:results}.
    The rendered images are zoomed in and white on black (instead of black on white) here and throughout the paper for visualization purposes.}
    \label{fig:cover}
\end{figure}

%% file: figure_2.tex
\begin{figure}
  \centering
  \begin{minipage}{0.49\linewidth}
    \begin{tcolorbox}[
      colback=white,
      colframe=orange!20,
      arc=1mm,
      boxsep=0pt,
      top=2pt,
      left=5pt,
      right=5pt,
      bottom=2pt,
      toptitle=3pt,
      bottomtitle=2pt,
      coltitle=black,
      title=\textbf{SynthPix}
    ]
      \includegraphics[trim={75 50 50 75},clip,width=0.49\linewidth]{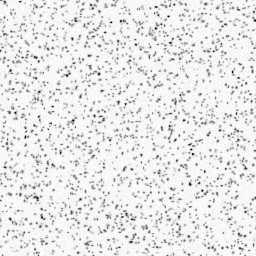}%
      \hfill
      \includegraphics[trim={150 50 50 150},clip,width=0.49\linewidth]{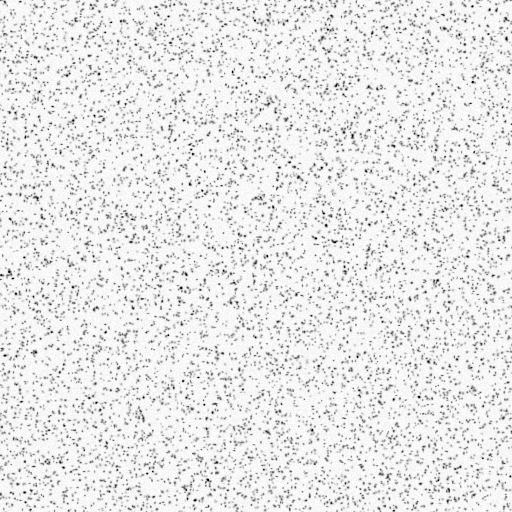}
    \end{tcolorbox}
  \end{minipage}
  \hfill
  \begin{minipage}{0.49\linewidth}
    \begin{tcolorbox}[
      colback=white,
      colframe=blue!20,
      arc=1mm,
      boxsep=0pt,
      top=2pt,
      left=5pt,
      right=5pt,
      bottom=2pt,
      toptitle=3pt,
      bottomtitle=3pt,
      coltitle=black,
      title=\textbf{PIVlab}
    ]
      \includegraphics[trim={150 50 50 150},clip,width=0.49\linewidth]{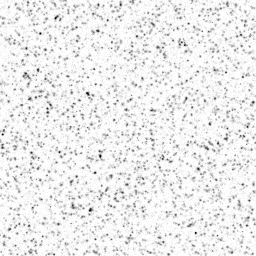}%
      \hfill
      \includegraphics[trim={300 50 50 300},clip,width=0.49\linewidth]{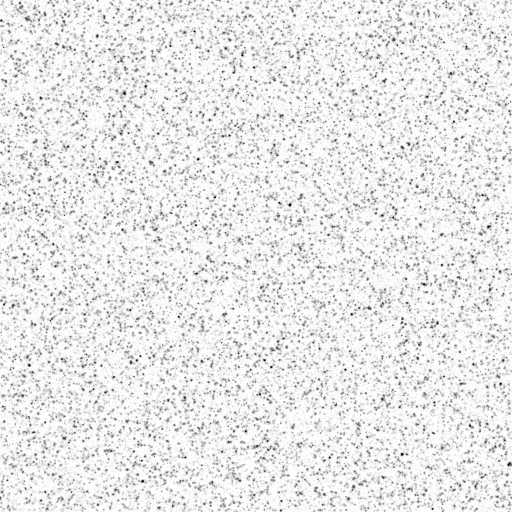}%
    \end{tcolorbox}
  \end{minipage}
    \caption{Zoomed-in images generated with SynthPix (left) and PIVlab \cite{stamhuis2014pivlab} (right). For each method we report images of resolution $256\times256$ and $512\times512$.}
    \label{fig:images-generated}
\end{figure}

%% file: figure_4.tex
\begin{figure}
    \centering
    \begin{minipage}{.49\linewidth}
    \begin{tcolorbox}[
      colback=white,
      colframe=red!20,
      arc=1mm,
      boxsep=0pt,
      top=2pt,
      left=5pt,
      right=5pt,
      bottom=2pt,
      toptitle=3pt,
      bottomtitle=3pt,
      coltitle=black,
      title=\textbf{Backstep}
    ]
    \begin{minipage}{.32\linewidth}
    \centering
    \scalebox{1}[-1]{
    \includegraphics[trim={1 1 10 1},clip,width=\linewidth]{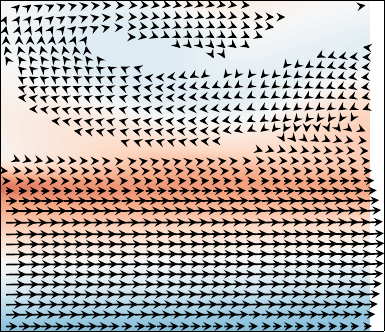}
    }
    \end{minipage}
    \hfill
    \begin{minipage}{.32\linewidth}
    \centering
    \includegraphics[%
    trim={100 90 90 100},clip,
    width=\linewidth]{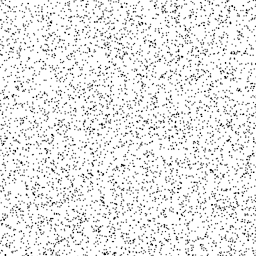}
    \end{minipage}
    \hfill
    \begin{minipage}{.32\linewidth}
    \centering
    \includegraphics[%
    trim={100 90 90 100},clip,width=\linewidth]{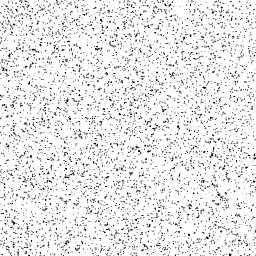}
    \end{minipage}
    \end{tcolorbox}
    \end{minipage}
    \hfill
    \begin{minipage}{.49\linewidth}
    \begin{tcolorbox}[
      colback=white,
      colframe=green!20,
      arc=1mm,
      boxsep=0pt,
      top=2pt,
      left=5pt,
      right=5pt,
      bottom=2pt,
      toptitle=3pt,
      bottomtitle=3pt,
      coltitle=black,
      title=\textbf{DNS}
    ]
    \begin{minipage}{.32\linewidth}
    \centering
    \scalebox{1}[-1]{
    \includegraphics[trim={10 1 10 1},clip,width=\linewidth]{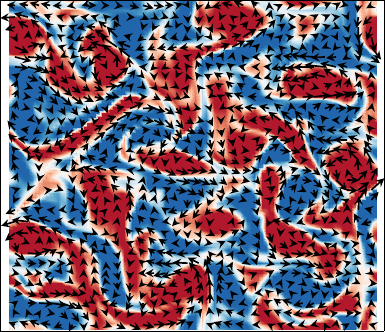}
    }
    \end{minipage}
    \hfill
    \begin{minipage}{.32\linewidth}
    \centering
    \includegraphics[%
    trim={150 60 60 150},clip,width=\linewidth]{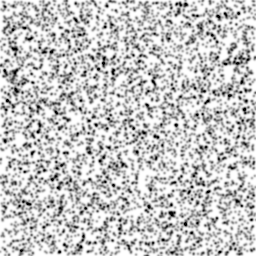}
    \end{minipage}
    \hfill
    \begin{minipage}{.32\linewidth}
    \centering
    \includegraphics[%
    trim={150 60 60 150},clip,width=\linewidth]{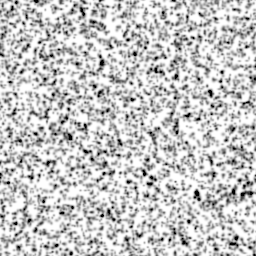}
    \end{minipage}
    \end{tcolorbox}
    \end{minipage}
    \caption{Flows (left), original (middle) and SynthPix-generated (right) images from two datasets of \cite{cai2019dense}.}
    \label{fig:accuracy}
\end{figure}

%% file: figure_5.tex
\begin{figure}
    \centering
    \begin{minipage}{\linewidth}
        \centering
        \begin{minipage}{.49\linewidth}
            \centering
            \includegraphics[width=\linewidth]{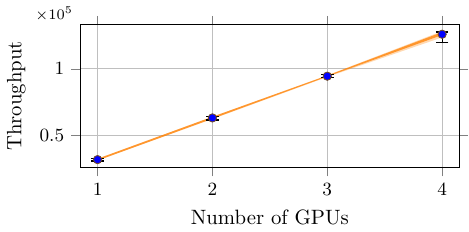}
        \end{minipage}
        \hfill
        \begin{minipage}{.49\linewidth}
            \centering
            \includegraphics[width=\linewidth]{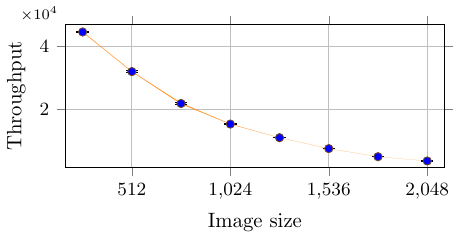}
        \end{minipage}
    \end{minipage}
    
    \begin{minipage}{\linewidth}
        \centering
        \begin{minipage}{.49\linewidth}
            \centering
            \includegraphics[width=\linewidth]{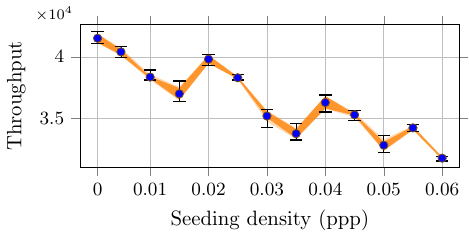}
        \end{minipage}
        \hfill
        \begin{minipage}{.49\linewidth}
            \centering
            \includegraphics[width=\linewidth]{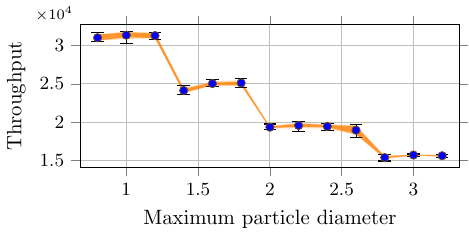}
        \end{minipage}
    \end{minipage}
    
    \begin{minipage}{\linewidth}
        \centering
        \begin{minipage}{.49\linewidth}
            \centering
            \includegraphics[width=\linewidth]{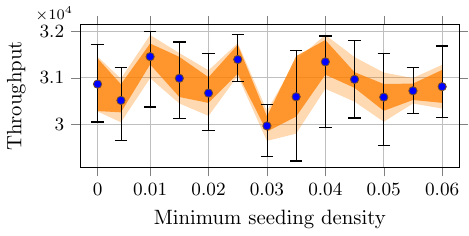}
        \end{minipage}
        \hfill
        \begin{minipage}{.49\linewidth}
            \centering
            \includegraphics[width=\linewidth]{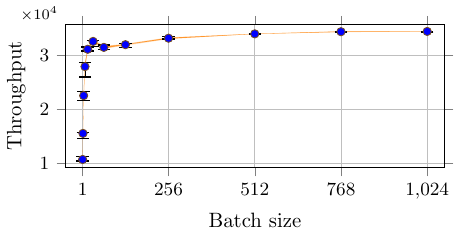}
        \end{minipage}
    \end{minipage}
    \begin{minipage}{\linewidth}
        \centering
        \begin{minipage}{.49\linewidth}
            \centering
            \includegraphics[width=\linewidth]{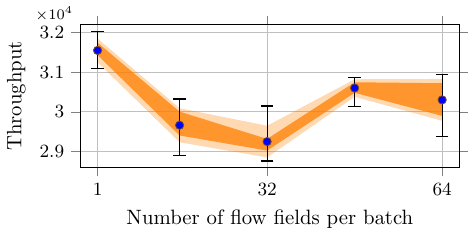}
        \end{minipage}
        \hfill
        \begin{minipage}{.49\linewidth}
            \centering
            \includegraphics[width=\linewidth]{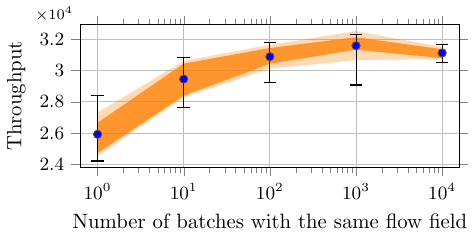}
        \end{minipage}
    \end{minipage}
    \caption{Results of the ablation studies in \cref{sec:results:scaling-laws}. Each plot shows how the throughput (in image pairs per second) changes with a single hyperparameter. The filled circles mark the mean over all the batches (we collect $1000$ batches for each hyperparameter). The black vertical error bars extend between the minimum and maximum observed values. The orange bands represent the mean $\pm$ one standard deviation (light) and the inter-quartile range (Q1–Q3, dark).}
    \label{fig:ablations-throughput}
\end{figure}